# Structural and magnetic properties of Mn-implanted Si

Shengqiang Zhou, K. Potzger, Gufei Zhang, A. Mücklich, F. Eichhorn, N. Schell, R. Grötzschel, B. Schmidt, W. Skorupa, M. Helm, and J. Fassbender

Institute for Ion Beam Physics and Materials Research, Forschungszentrum Rossendorf, POB 510119, 01314 Dresden, Germany

D. Geiger

Institute of Structure Physics, Dresden University, Zellescher Weg 16, 01062 Dresden, Germany

Abstract

Structural and ferromagnetic properties in Mn implanted, p-type Si were investigated. High resolution structural analysis techniques like synchrotron X-ray diffraction revealed the formation of $MnSi_{1.7}$ nanoparticles already in the as implanted samples. Depending on the Mn-fluence, the size increases from 5 nm to 20 nm upon rapid thermal annealing. No significant evidence is found for Mn substituting Si sites either in the as-implanted or annealed samples. The observed ferromagnetism yields a saturation moment of 0.21 $\mu_B$ per implanted Mn at 10 K, which could be assigned to $MnSi_{1.7}$ nanoparticles as revealed by a temperature dependent magnetization measurement.

# I. Introduction

Recently, diluted magnetic semiconductors (DMS) have attracted huge research attention [1, 2]. III-V and II-IV compound semiconductors doped with transition metals (TM) by ion-implantation show ferromagnetism with a Curie temperature up to room temperature [3, 4]. However, Si-based DMS would be preferable to compound semiconductors due to the availability of high quality Si in large sizes at relatively low costs. More importantly, the fabrication of Si-based DMS is compatible with the mature microelectronics technique. Based on the Zener model, Dietl *et al.* [5] predicted a carrier-mediated ferromagnet if Si is doped with 5% Mn. Zhang *et al.* [6] reported on the crystalline $Mn_{0.05}Si_{0.95}$ alloy with a Curie temperature up to 400 K. Also epitaxially grown $Mn_{0.07}Si_{0.93}$ thin films show ferromagnetic ordering up to 200 K, and at the same time, possess semiconducting properties [7]. In addition to the aforementioned epitaxial techniques, ion-implantation is another promising method for ferromagnetic doping of semiconductors, e.g., implantation of almost any TM element with an atomic concentration exceeding those possible in thermal equilibrium by far. Very recently, Bolduc *et al.* [8] have reported room temperature ferromagnetism of Mn-ion implantation Si at Mn fluences of $1\times10^{15}\,cm^{-2}$ and $1\times10^{16}\,cm^{-2}$. The Curie temperature has been evaluated to be above 400 K for both n-type and p-type Si. The strength of the ferromagnetism was affected by the Mn concentration, thermal annealing and carrier type of the Si substrate. The ferromagnetic coupling was believed to be carrier-mediated. However, the high-resolution structural characterization of the investigated Si:Mn systems (both epitaxially grown and ion-implanted) was rather limited. The diffusion of Mn in Si upon annealing was only investigated by Rutherford backscattering (RBS) and secondary ion mass spectrometry [9, 10]. Generally Mn silicides are easily formed. E.g. if Mn is deposited on Si, an epitaxial MnSi is formed [11-13]. Also transition metal (i.e. Co and Ni) implantation into Si can easily leads to silicides [14, 15]. Therefore a careful characterization of the Mn doped Si samples has to be carried out in order to clarify the origin of the ferromagnetism, i.e. the formation of nanoparticles or a DMS system. For this purpose, in the present paper, RBS/channeling, transmission electron microscopy (TEM) and synchrotron radiation x-ray diffraction (SR-XRD) are used to clarify the possible origin of the ferromagnetism in Mn-implanted Si.

## II. Experiment

The Mn-implanted Si samples were prepared from commercially available, Czochralski grown single-crystal Si(001) wafers, which were p-type doped with a B concentration of $1.2 \times 10^{19}$ cm$^{-3}$. Mn$^+$ ions were implanted at an energy of 300 keV with fluences of $1 \times 10^{15}$ cm$^{-2}$, $1 \times 10^{16}$ cm$^{-2}$ and $5 \times 10^{16}$ cm$^{-2}$, which corresponds to Mn peak concentrations of 0.08, 0.8 and 4 at %, respectively, with a projected range ($R_p$) of 258±82 nm. The samples were held at 350 °C during implantation to avoid amorphization. In order to reduce channeling effects, the angle between the sample surface normal and the incident beam was set to 7°. After implantation, rapid thermal annealing (RTA) was performed at a temperature of 800 °C for 5 min in a forming gas of $N_2$. The RBS spectra were taken with a collimated 1.7 MeV He$^+$ beam with a surface barrier detector at 170°. TEM specimens were conventionally prepared for cross-section inspection and examined on a Philips CM 30 FEG microscope. Conventional XRD (Conv. XRD) measurements were performed on a Siemens D5005 diffractometer using Cu irradiation. SR-XRD was performed at Rossendorf beamline (ROBL) at European Synchrotron Radiation Facility with monochromatic X-rays of 0.1541 nm wavelength. Magnetic properties were analyzed using a superconducting quantum interference device (SQUID) magnetometer (Quantum Design MPMS).

# III. Results and discussion

## A. Mn lattice location

After Mn-ion implantation, the samples were analyzed by RBS/channeling to check the implantation induced damage and the lattice location of Mn with respect to the Si lattice. Figure 1 shows the RBS/channeling spectra for all samples. The Mn signals display a Gaussian distribution which corresponds to an $R_p$ of around 258 nm. For samples with a fluence of $1\times10^{16}$ cm$^{-2}$ and $5\times10^{16}$ cm$^{-2}$, the $R_p$ and the amount of Mn agree well with SRIM simulation [16] and designed values. However, for the lowest fluence sample ($1\times10^{15}$ cm$^{-2}$), the concentration of Mn is in the order of the detection limit. All $\chi_{min}$ values, the ratio of the backscattering yield at channelling condition to that for a random beam incidence [17], which label the degree of lattice disordering, for Si and Mn signals are listed in Table I. As expected, a higher fluence implantation introduces a larger lattice disordering (a larger $\chi_{min}$) in the silicon substrates. For the Mn signal, $\chi_{min}$ amounts to more than 90% for both samples with the fluence of $1\times10^{16}$ cm$^{-2}$ and $5\times10^{16}$ cm$^{-2}$, which proves that the majority of the implanted Mn atoms are not substituted to Si sites with respect to the [001] crystallographic axis.

Figure 2 shows the RBS spectra for the sample implanted with $5\times10^{16}$ cm$^{-2}$ Mn ions after RTA annealing at 800 °C for 5 min. The changes of $\chi_{min}$ values are listed in Table I. The annealing process removed only a very small fraction of the implantation induced damage (a slight decrease of $\chi_{min}$) and did not change the Mn element distribution, i.e. Mn atoms did not diffuse to deep or shallow places, nor to substitutional Si sites. The other samples show similar behavior after annealing (see Table I).

## B. Observation of crystalline MnSi$_x$ nanoparticles

In order to confirm the formation of Mn-silicides nanoparticles, high resolution TEM (HRTEM) was performed. For the highest fluence ($5\times10^{16}$ cm$^{-2}$), some small nanoparticles with diameters of ~5 nm are detected already in the as-implanted samples (not shown). Figure 3 shows typical TEM-overview images at lower magnification of samples after RTA treatment. Nano-sized precipitates are unambiguously found in all annealed samples with different fluence. They exhibit mean diameters of 5 nm for a fluence of $1\times10^{15}$ cm$^{-2}$, 10 nm for a fluence of $1\times10^{16}$ cm$^{-2}$, and 20 nm for a fluence of $5\times10^{16}$ cm$^{-2}$. Figs. 4(a) and (b) show HRTEM images for the samples of 1E16 RTA and 5E16 RTA (see Table I for the sample abbrevation). Most of those nanoparticles are spherical-like, their diameters increase from around 10 nm to 20 nm with increasing fluence. However, considering the rich varieties of Mn-silicides [13], it is extremely difficult to identify these nanoparticles even by HRTEM [18]. Table II lists the structural information of some higher manganese silicides with Si composition of around 1.7 after Ref. 13. All different compositions are named as MnSi$_{1.7}$ throughout the rest of the article. The crystallite area of some particles was Fourier-analysed as shown in Figure 4(c). The reflections in the diffractogram indicated by arrows reveal a lattice spacing of around 1.7 nm, which can be attributed to MnSi$_{1.7}$ (see Table II).

In addition, XRD was performed in order to confirm the formation of Mn-silicide nanoparticles. Since a conventional XRD diffractometer (Siemens D5005) failed to detect any crystalline nanoparticles in both as-implanted and annealed samples ($5\times10^{16}$ cm$^{-2}$) due to the small amount of nanoparticles and the limited intensity of Cu-target x-ray source, we performed XRD measurements at a synchrotron radiation source. However, also at synchrotron XRD, in a symmetric beam geometry (Bragg-Brentano) one fails to detect any Mn-silicides in all samples (not shown). Therefore, a grazing incidence geometry was used during the measurement. Grazing incidence X-ray diffraction (GIXRD) refers to a method where the incident X-ray beam is aligned at a small (here 0.4°) angle to the sample surface. This has the advantage to limit the penetration depth of the X-rays into the sample with consequently low background scattering from the substrate. Figure 5 shows GIXRD patterns of all samples. The diffraction peaks at around 42° and 46.3° cannot be attributed to the Si substrate, but to $MnSi_{1.7}$ (see Table II). For the highest fluence sample ($5\times10^{16}$ cm$^{-2}$), nanoparticles were observed already in the as-implanted state, and they grew from around 5 nm to 15 nm in grain size after annealing. $MnSi_{1.7}$ was also found for a fluence of $1\times10^{16}$ cm$^{-2}$ after annealing, while there is no indication of any crystalline nanoparticles in the lowest fluence sample ($1\times10^{15}$ cm$^{-2}$) even after annealing. Using Scherrer formula [19], the average size of these nanoparticles is calculated, and listed in Table I.

## C. Magnetic properties

Figure 6 shows the magnetization versus field curves (M-H loops) recorded at 10 K. The sample of 1E16 RTA shows a clear hysteretic behavior with a saturation of 0.21 $\mu_B$/Mn and a coercivity of 275± 25 Oe. In contrast, both samples of 5E16 asimp. and 5E16 RTA show very weak hysteretic behaviors. All other samples (1E16 Asimp, 1E15 Asimp. and 1E15 RTA) did not show any detectable magnetism (not shown here). In order to clarify the origin of the ferromagnetism of the sample 1E16 RTA, zero-field cooled (ZFC) and field cooled (FC) magnetization curves were measured in a field of 5 mT from 10 to 300 K (the inset of Figure 6). The ZFC curve shows a gradual increase at low temperature, and reaches a maximum at a temperature of $T_B$ (around 23 K), while the FC curve continues to decrease with increasing temperature. At higher temperature than $T_B$, FC curves merges together with ZFC curve. Figure 6 also shows a M-H loop taken at 100 K. Below $T_B$ (10 K), these nanoparticles show ferromagnetic behavior, while above $T_B$ (100 K), the M-H loop shows neither cleary remanence nor coercivity and these nanoparticles are superparamagnetic. The above-mentioned magnetic behavior is typical for systems containing magnetic nanoparticles [20-22]. Moreover, the wasp-waist shape of the loop is associated with magnetic phases with different coercivities, which could be attributed to the size distribution of nanoparticles as indicated by HRTEM results [23-25]. Although the sample preparation in our work is identical with Ref. 8, the magnetic properties are drastically different. There could be some other parameters, or technical tricks in sample preparation to result in the discrepancy.

## D. Discussion

In Table I, all structural and magnetic properties of the investigated samples are listed. Obviously the dependence of magnetization on the fluence and annealing is rather anomalous. In order to explain the magnetization, the magnetic properties for Mn-silicides have to be reviewed. Unfortunately very limited work has been done on the magnetism of $MnSi_{1.7}$ [26]. In Ref. 26, single crystalline $Mn_4Si_7$ ($MnSi_{1.75}$) were reported to exhibit weak itinerant magnetism with an ordering temperature of 47 K and with a very low saturation moment of 0.012 $\mu_B$/Mn. For the nanoparticles embedded in Si, i.e. with different surface/volume ratio and undergoing different pressure, the magnetism can be much different from bulk crystals. During intense investigations on MnSi (a weak itinerant magnetic material), the ordering temperature has been found to strongly depend on lattice strain, induced, e.g., by hydrostatic pressure or by the lattice mismatch with substrate [27-30]. If the exchange interaction is changed for instance by pressure or by alloying, the system no longer orders even at T=0 K [27]. Considering the different grain size of MnSi$_{1.7}$ in our sample, i.e. different stress provided by the surrounding lattice, we could explain the dependence of the magnetization on the Mn fluence and post-implantation annealing as following. Those MnSi$_{1.7}$ nanoparticles in the sample (1E16 RTA) have a particular medium grain size, i.e. particular lattice deformation or strain, therefore a higher ordering temperature and even a much higher moment (0.21 $\mu_B$/Mn) than the bulk crystal (0.012 $\mu_B$/Mn). While the bigger nanoparticles in the sample of 5E16 RTA are more bulk-like and have very weak ferromagnetism. Those smaller MnSi$_{1.7}$ nanoparticles in other samples (5E16 Asimp. and 1E15 RTA) could be in a rather worse crystalline quality or different strain status, and therefore only show weak or non-detectable ferromagnetism. However a non-ambiguous reason for the anomalous dependence of the magnetization on the particle size, however, is not yet known.

## IV. Conclusion

By investigating Mn-ion implanted Si, we have found a non-systematic dependence of the induced ferromagnetization on the implantation fluence. No significant indication for Mn substituting Si either in the as-implanted or the rapidly annealed samples. The observed ferromagnetism is attributed to MnSi$_{1.7}$ nanoparticles. The magnetization is maximized at a certain medium grain size of MnSi$_{1.7}$ nanoparticles, which could result in a particular strain status, consequently a stronger ferromagnetic coupling. Moreover high resolution analysis techniques are necessary in order to identify those nanoparticles and to clarify the origin of the observed ferromagnetism.

Acknowledgement

We thank Dr. S. Gemming for valuable discussion.

Table I: Map of structural and magnetic properties of all samples

| Sample | 1E15 Asimp. | 1E15 RTA | 1E16 Asimp. | 1E16 RTA | 5E16 Asimp. | 5E16 RTA |
|---|---|---|---|---|---|---|
| Fluence and process | $1\times10^{15}$ cm$^{-2}$, As-implanted | $1\times10^{15}$ cm$^{-2}$, RTA[1] | $1\times10^{16}$ cm$^{-2}$, As-implanted | $1\times10^{16}$ cm$^{-2}$, RTA[1] | $5\times10^{16}$ cm$^{-2}$, As-implanted | $5\times10^{16}$ cm$^{-2}$, RTA[1] |
| $\chi_{min}$-Si (RBS/C) | 7.4% | 6.4% | 51% | 41% | 82% | 70% |
| $\chi_{min}$-Mn (RBS/C)[2] | - | - | 91% | 89% | 96% | 98% |
| Conv. XRD[3] | No | No | No | No | No | No |
| TEM[3] | No | Yes | No | Yes | Yes | Yes |
| SR-XRD[3] | No | No | No | Yes (11 nm[4]) | Yes (6 nm[4]) | Yes (15 nm[4]) |
| Ferromagnetism (10K) | No | No | No | Strong (0.21$\mu_B$/Mn) | Weak (0.04$\mu_B$/Mn) | Weak (0.02$\mu_B$/Mn) |

[1] RTA: rapid thermal annealing in a forming gas of N$_2$ at 800 °C for 5 min.

[2] For the Mn signal, a very weak channeling effect was observed, which reveals that the majority of implanted Mn ions are not substituted on Si sites.

[3] Yes or No to answer if precipitates of Mn-silicides are detectable by corresponding technique.

[4] Average grain size.

Table II Composition and lattice parameters of tetragonal MnSi$_{1.7}$ phases (after ref. 13).

| Phase | Composition | a (nm) | c (nm) | Space group | PDF number[1] |
|---|---|---|---|---|---|
| Mn$_4$Si$_7$ | 1.750 | 0.5525 | 1.7463 | P-4c2 | 72-2069 |
| Mn$_{11}$Si$_{19}$ | 1.727 | 0.5518 | 4.8136 | P-4n2 | - |
| Mn$_{15}$Si$_{26}$ | 1.733 | 0.5525 | 6.5550 | I-42d | 20-724 |
| Mn$_{27}$Si$_{47}$ | 1.741 | 0.5530 | 11.7940 | P-4n2 | 26-1251 |

[1] Inorganic Crystal Structure Database (ICSD)

Fig captions

Fig. 1. RBS random and channeling along Si[001] spectra of as-implanted Si samples. Random (○), channeling (+) for the fluence of $5\times10^{16}$ cm$^{-2}$; channeling (■) for the fluence of $1\times10^{16}$ cm$^{-2}$; channeling (Δ) for the fluence of $1\times10^{15}$ cm$^{-2}$; and channeling for virgin Si (—) which is slightly below triangle symbols.

Fig. 2. RBS random and channeling along Si[001] spectra for sample implanted with a fluence of $5\times10^{16}$ cm$^{-2}$. Random (○) and channeling (+) for as-implanted sample, channeling (—) for RTA sample and channeling (Δ) for virgin Si. Only a small fraction of damage was recovered by annealing. For Mn signals, no channeling was observed either for as-implanted or annealed samples, which means Mn does not substitute Si lattice site.

Fig. 3. TEM-overview images show the implantation induced damage and nano-sized precipitates after RTA: (a) $1\times10^{15}$ cm$^{-2}$, (b) $1\times10^{16}$ cm$^{-2}$ and (c) $5\times10^{16}$ cm$^{-2}$. The nano-sized precipitates are visible as spot-like contrast, and they are growing with increasing fluence. The arrow marks the direction towards the sample surface.

Fig. 4. High resolution TEM images for representative MnSi$_x$ nanoparticles in (a) 1E16 RTA, (b) 5E16 RTA, and (c) the diffractogram of one nanoparticle in (b). The reflection spots marked by the arrows in (c) reveal a lattice spacing of around 1.7 nm, which can be attributed to MnSi$_{1.7}$.

Fig. 5. XRD grazing incidence scans of all investigated samples implanted with Mn. In some samples two peaks arise, which can be attributed to Mn-silicide nanoparticles (MnSi$_{1.7}$).

Fig. 6 M-H hysteresis loops for Mn-implanted Si samples. Only the sample of 1E16 RTA exhibits a clear hysteretic behavior. Inset: ZFC (open circle) and FC (solid line) magnetization curves reveal a typical characteristic of a magnetic nanoparticle system for the sample of 1E16 RTA. Above the blocking temperature, the sample of 1E16 RTA shows superparamagnetic properties, where neither coercivity nor remanence was observed at 100 K. None of other samples (1E15 RTA, 1E15 Asimp, and 1E16 Asimp) exhibit ferromagnetism even at temperatures down to 10 K (not shown).

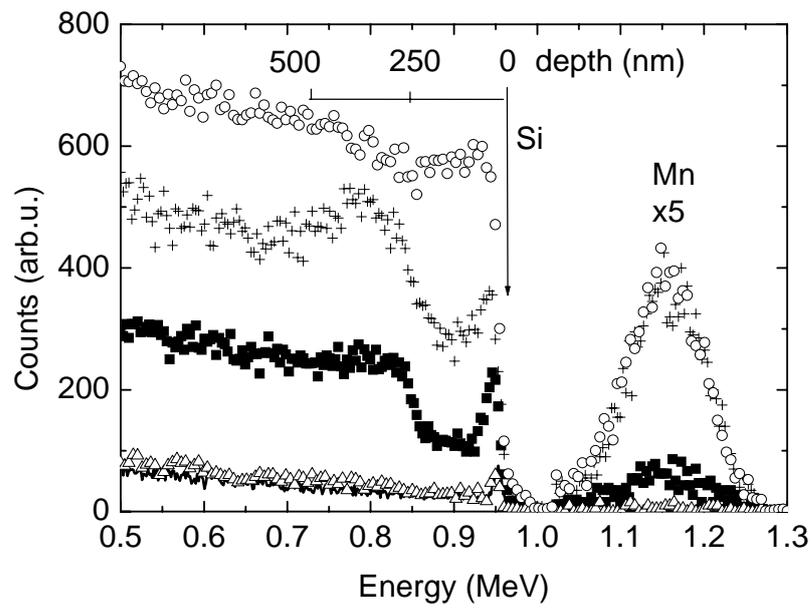

Fig. 1.

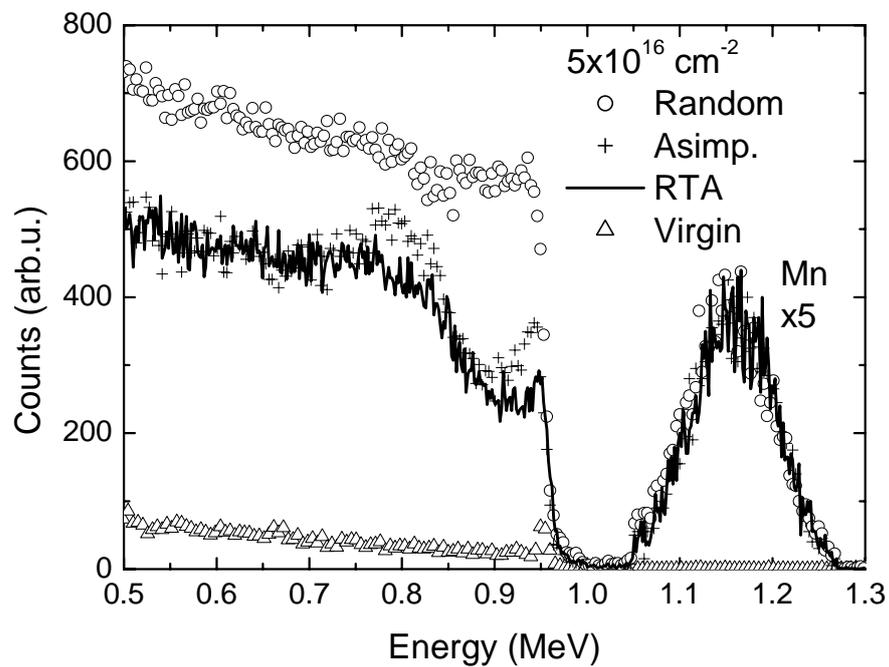

Fig. 2.

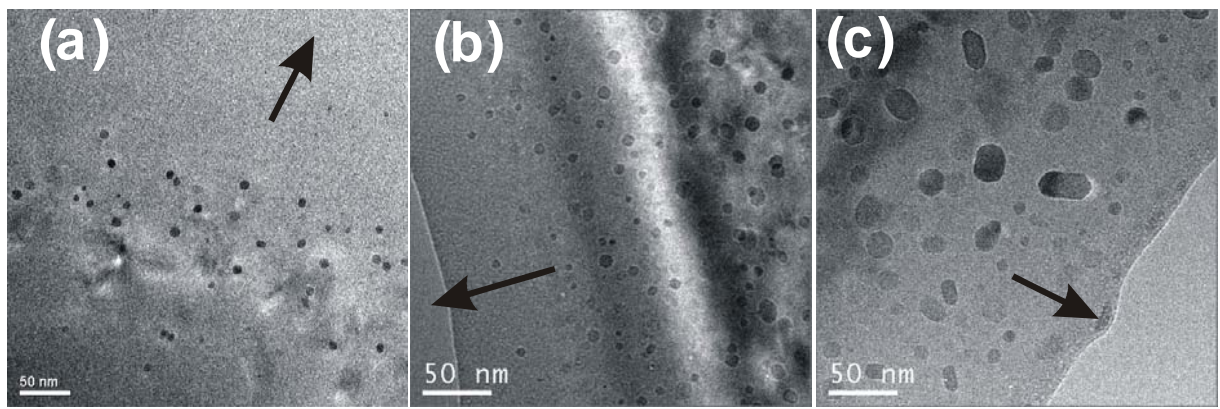

Fig. 3.

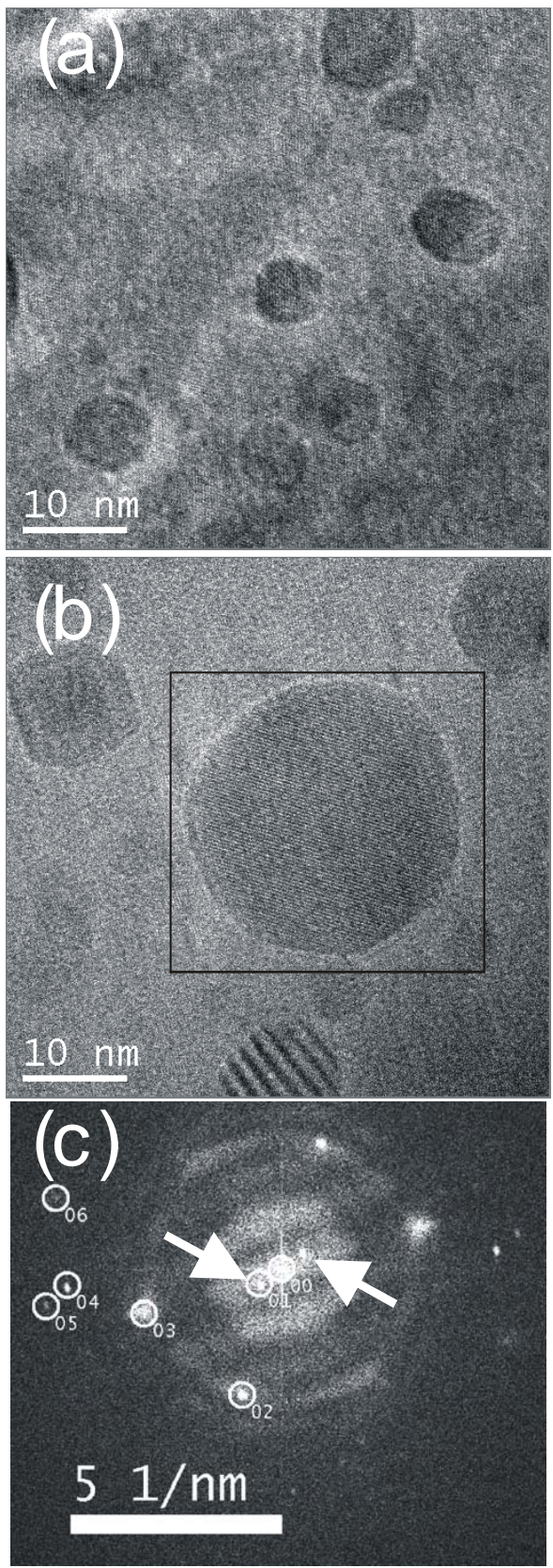

Fig. 4.

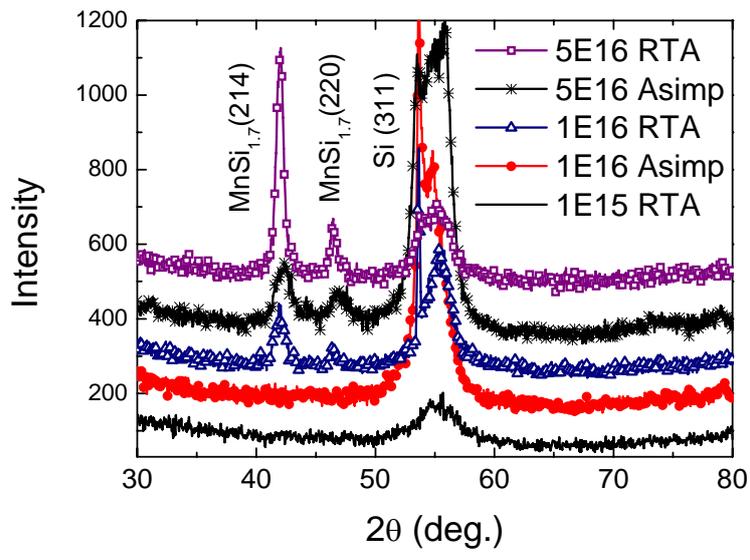

Fig. 5.

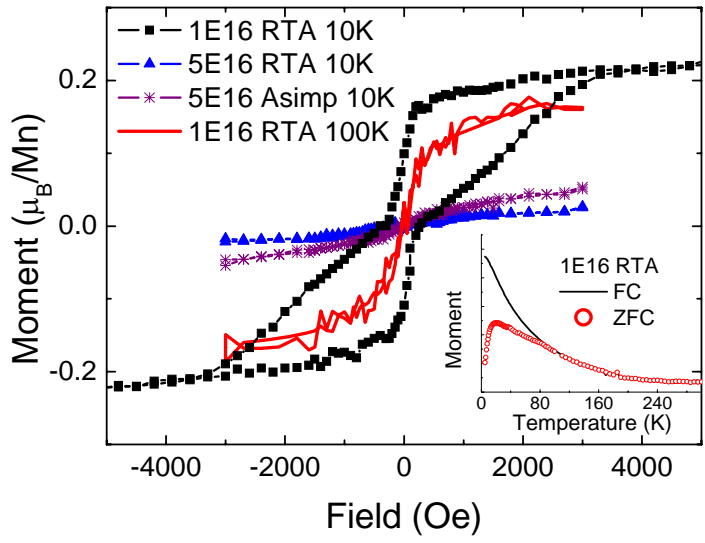

Fig. 6